\documentclass[preprint,12pt]{elsarticle}
\usepackage{amssymb}
\usepackage{amsmath}

\usepackage{lineno}

\journal{Nuclear Instruments and Methods in Physics Research A}

\begin{document}

\begin{frontmatter}

\title{Analytical Fitting of $\gamma$--ray Photopeaks in Germanium Cross Strip Detectors}

\author[a,b]{Steven E. Boggs\corref{cor1}}
\ead{seboggs@ucsd.edu}
\cortext[cor1]{Corresponding author.}

\author[a]{Sean N. Pike}

\affiliation[a]{organization={Department of Astronomy \& Astrophysics, University of California, San Diego},
            addressline={9500 Gilman Drive}, 
            city={La Jolla},
            state={CA},
            postcode={92093}, 
            country={USA}}
            
\affiliation[b]{organization={Space Sciences Laboratory, University of California, Berkeley},
            addressline={7 Gauss Way}, 
            city={Berkeley},
            state={CA},
            postcode={94720}, 
            country={USA}}

\begin{abstract}

In an ideal germanium detector, fully-absorbed monoenergetic $\gamma$--rays will appear in the measured spectrum as a narrow peak, broadened into a Gaussian of width determined only by the statistical properties of charge cloud generation and the electronic noise of the readout electronics. Multielectrode detectors complicate this picture. Broadening of the charge clouds as they drift through the detector will lead to charge sharing between neighboring electrodes and, inevitably, low-energy tails on the photopeak spectra. We simulate charge sharing in our germanium cross strip detectors in order to reproduce the low-energy tails due to charge sharing. Our goal is to utilize these simulated spectra to develop an analytical fit (shape function) for the spectral lines that provides a robust and high-quality fit to the spectral profile, reliably reproduces the interaction energy, noise width, and the number of counts in both the true photopeak and the low-energy tail, and minimizes the number of additional parameters. Accurate modeling of the detailed line profiles is crucial for both calibration of the detectors as well as scientific interpretation of measured spectra.

\end{abstract}



\end{frontmatter}


\section{Introduction}
\label{sect:intro}

The Compton Spectrometer and Imager (COSI) is a soft $\gamma$--ray survey telescope (0.2-5\,MeV) designed to probe the origins of Galactic positrons, reveal sites of ongoing element formation in the Galaxy, use $\gamma$--ray polarimetry to gain insight into extreme environments, and explore the physics of multi-messenger events \cite{kierans20172016,sleator2019benchmarking,arxiv.2109.10403}. The COSI detectors are custom, large-volume (54\,cm$^2$ area, 1.5\,cm thick) cross-strip germanium detectors utilizing amorphous contact technologies \cite{amman2007amorphous}. Cross-strip electrodes on the opposite faces, combined with signal timing, provide full 3D position resolution for interactions within the detector. In this work we are focused on our original 2.0-mm strip pitch germanium detectors that flew on the COSI balloon payload \cite{kierans20172016,sleator2019benchmarking}.

When a $\gamma$--ray photon interacts in the germanium, either by photoabsorption or Compton scattering, a fast recoil electron is produced which knocks more electrons from the valence band to the conduction band, leaving holes behind. The number of electron-hole (e-h) pairs is directly proportional to the energy deposited, 2.96 eV per e-h pair in germanium. In an applied electric field (+1500 V bias) these charge clouds will separate and drift in opposite directions, electrons toward the cathode and holes toward the anode. The $\gamma$--ray interaction energy is measured on both the cathode (electron signal) and the anode (hole signal) strips independently by measuring the integrated charge induced on each electrode by the charge clouds drifting towards their respective electrodes. As these charge clouds drift in the detector, their charge density profiles broaden due to both thermal diffusion and mutual electrostatic repulsion. The finite size of the charge clouds will result in some interactions having their charge collected on multiple electrodes. Such interactions lead to either charge sharing between strips, or low-energy tailing on spectral lines if the charge shared on the neighboring strip falls below the detection threshold for that electrode. 

Optimizing the spectral performance of these high-resolution germanium detectors requires detailed knowledge of the photopeak and low-energy tail profiles. Inadequate modeling of the shapes of the spectral line can affect the overall spectral calibration of the instrument, as well as scientific analysis of observed spectral features. 

\begin{figure}
\centering
\includegraphics[width=0.8\textwidth]{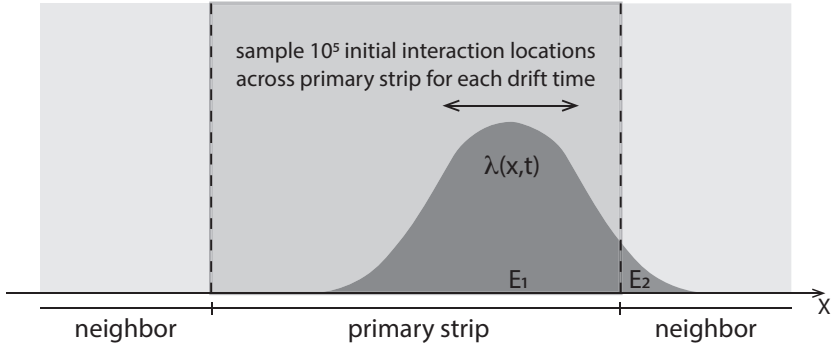}
\caption{\label{fig:f1} Diagram of the process utilized to create the simulated spectra, including the true photopeak, low-energy tail, and ``measured'' total. For a given drift time, $\tau$, we sampled $10^5$ initial interaction positions across the primary strip. The resulting charge cloud profiles, $\lambda (x,\tau)$, were numerically integrated to determine the charge (energy), $E_1$, collected on the primary strip as well as that collected on the neighbor, $E_2$.}
\end{figure}

We present a novel shape function that reliably reproduces both the line profiles and the underlying physical parameters for our spectra created with charge sharing simulations, with minimal additional fit parameters.

In Section~\ref{sect:spectra}, we describe how we create simulated photopeak and low-energy tail spectra utilizing a novel charge profile density model. In Section~\ref{sect:previous} we review how germanium spectral lines with low-energy tails have been fit in the past. Section~\ref{sect:tail-fit} presents the shape function utilized in this work.  In Section~\ref{sect:constrain} we show how constraining some of the shape function parameters leads to more robust estimates of the underlying physical parameters.  Section~\ref{sect:general} extends these fits to higher and lower interaction energies. In Section~\ref{sect:extended} we demonstrate how the shape function varies for line profiles that include the effects of extended initial charge clouds (created by the recoil electron) and present fit parameters accounting for effects of the recoil electrons. We conclude with a discussion of applications and future directions.

\section{Model Spectra}
\label{sect:spectra}

\begin{figure}
\centering
\includegraphics[width=1.0\textwidth]{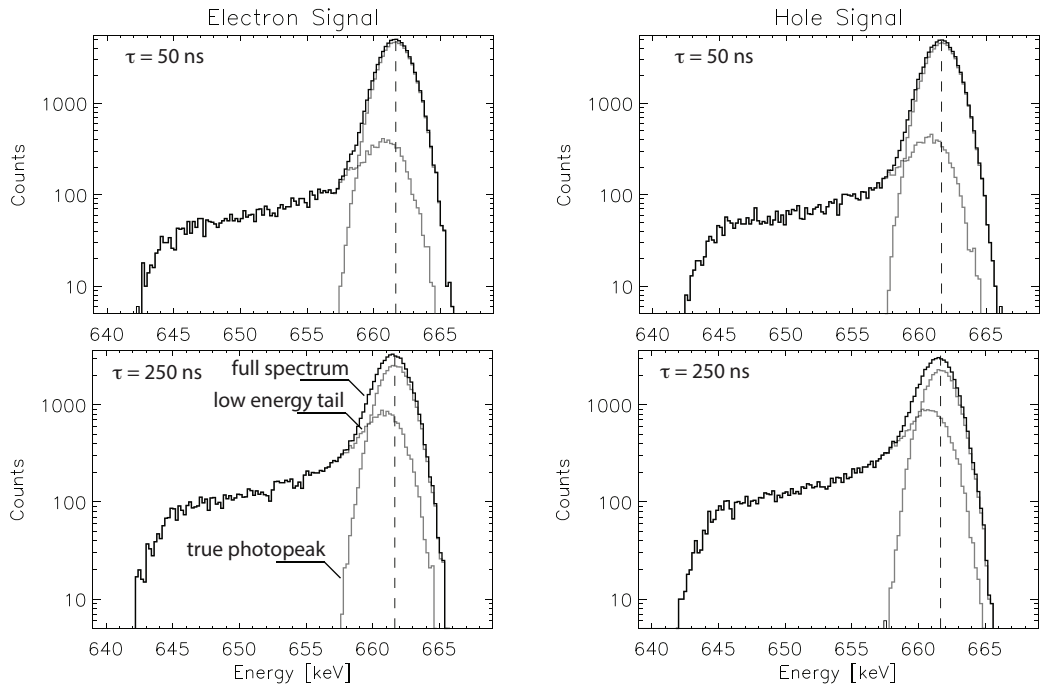}
\caption{\label{fig:f2} Example model spectra for 661.66\,keV photopeak interactions, showing the individual true photopeak and low-energy tail components of the model as well as the combined full spectra. (Top Left) Electron signals, $\tau = 50\,ns$ drift time, (Top Right) hole signals, $\tau = 50\,ns$, (Bottom Left) electron signals, $\tau = 250\,ns$, (Bottom Right) hole signals, $\tau = 250\,ns$. Events with $\tau = 50\,ns$ occur near the signal collection electrode, while those with $\tau = 250\,ns$ occur far from the collection electrode.}
\end{figure}

We utilize the analytical charge cloud profiles derived in \citet{boggs2023} to simulate charge sharing within our germanium cross strip detectors. The analytical approximations in that paper allow us to model the 1-D projected charge density profiles for electron and hole clouds across the collection electrodes as a function of their drift time ($\tau$) in the detector, which maps directly to interaction depth within the detector. These charge profiles include the effects of thermal diffusion and mutual electrostatic repulsion in broadening the charge clouds. In this work we initially assume that the recoil electron depostis all of its energy at a single point, but we include the effects of a finite initial charge cloud distributions in Section~\ref{sect:extended}. In order to turn these charge density profiles for fixed drift times into spectra, we sampled 10$^5$ initial interaction locations across the primary strip electrode for each drift time (Fig.~\ref{fig:f1}), numerically integrating the charge density profile to determine the charge (energy) deposited on the primary strip ($E_1$) and neighboring strip ($E_2$). Once $E_1$ was determined for each sampled location we added a random noise to $E_1$ based on our measured resolution $\sigma_m(E)$, which is given approximately by:

\begin{equation}
\sigma_m(E) = [2.17+0.65*\sqrt{E/1000}]/(2.35)\,keV; [E] = keV
\label{eqn:noise}
\end{equation}

Each of these simulated interactions was classified into one of three categories based on the charge (energy) deposited on the neighboring strip. Events where $E_2 = 0$ (no charge sharing) were classified as ``true photopeak'' events and contribute to the narrow Gaussian photopeak centered at the initial interaction energy, $E_0$. Events where enough charge was shared on the neighboring strip to exceed the trigger threshold of the readout electronics on that strip ($E_2 \ge E_{th}$) were classified as ``triggered shared'' events. We are focused on analytical fitting of the single-strip photopeak spectra in this work and hence are not considering these triggered shared events any further here. The last classification is for events where charge (energy) was collected on the neighboring strip, but not enough to trigger the strip ($0 < E_2 < E_{th}$). These events were classified as ``untriggered shared'' events and contribute to the low-energy tail on the spectral peak. (The COSI readout electronics that flew on the balloon payload, which we are modeling in this work, were not designed to read out neighboring strips unless the interaction on the neighbor exceeded the trigger threshold.) We then proceed to bin the ``measured'' energy $E_1$ into one of two spectra, the true photopeak spectrum and the low-energy tail spectrum. Separating these two spectral components through the modeling allows us to know the exact number of counts in the true photopeak ($N_{p}$) and the low-energy tail ($N_{t}$) separately, as well as study the shape of the low-energy tail independently of the true photopeak. Adding these two spectra together creates our ``measured'' full spectral line for a given drift time.

\begin{table}
\centering
\begin{tabular}{ccccc}
\hline
Electron signals\\
\hline
$\tau$ [ns] & $E_{0}$ [$keV$] & $\sigma$ [$keV$] & $N [fit/sim]$ & $\chi_{R}^{2}$\\
\hline
10 & 661.62 & 1.18 & 0.97 & 21.22\\
50 & 661.56 & 1.21 & 0.94 & 40.94\\
100 & 661.51 & 1.23 & 0.90 & 55.60\\
150 & 661.46 & 1.26 & 0.88 & 66.34\\
200 & 661.41 & 1.28 & 0.85 & 75.19\\
250 & 661.34 & 1.30 & 0.83 & 82.52\\
\hline
& & & \\

\hline
Hole signals\\
\hline
$\tau$ [ns] & $E_{0}$ [$keV$] & $\sigma$ [$keV$] & $N [fit/sim]$ & $\chi_{R}^{2}$\\
\hline
10 & 661.61 & 1.18 & 0.97 & 20.90\\
50 & 661.55 & 1.22 & 0.93 & 43.55\\
100 & 661.49 & 1.25 & 0.90 & 59.44\\
150 & 661.42 & 1.27 & 0.87 & 71.38\\
200 & 661.36 & 1.30 & 0.84 & 80.38\\
250 & 661.28 & 1.33 & 0.81 & 88.02\\
\hline

\end{tabular}
\caption{\label{tab:662gaussian} Gaussian fit parameters for full spectra. The simulated spectral model assumes $E_{0} = 661.66\,keV$ and $\sigma = 1.15\,keV$. The quality of fits are poor ($\chi_{R}^{2} \gg 1$), even for the shortest drift times corresponding to minimal tailing. At longer drift times, the quality of fit degrades, and the fit values for $E_{0}$, $\sigma$, and the number of events in the peak ($N[fit/sim]$) become less accurate.}

\end{table}

\begin{figure}
\centering
\includegraphics[width=1.0\textwidth]{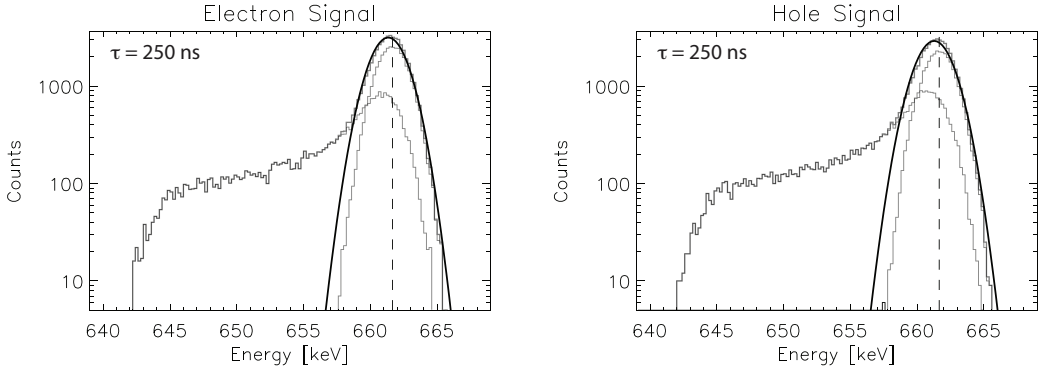}
\caption{\label{fig:f3} Gaussian-only fits to the full spectra, 250\,ns drift time. (Left) Electron signals, both the fitted (and expected) values, $E_0$ = 661.34 keV (661.66), $\sigma = 1.30\,keV$ (1.15). (Right) Holes signals, $E_0$ = 661.28 keV (661.66), $\sigma = 1.33\,keV$ (1.15). The individual true photopeak and low-energy tail components of the model spectra are shown for comparison. More details are presented in Table~\ref{tab:662gaussian}.}
\end{figure}

In Figure~\ref{fig:f2} we show example electron-signal and hole-signal spectra for two different drift times, $\tau = 50\,ns$ which represent interactions near the collection electrode and hence have minimal charge sharing with neighbor strips, and $\tau = 250\,ns$ which represents interactions far from the collection electrode so exhibit maximum charge sharing with neighbor strips. These spectra demonstrate how the untriggered shared events create the low-energy tails on the photopeaks in our simulated spectra. (For clarity, electron and hole spectra at the same drift time as shown here do not correspond to the same interactions as longer electron drift times would correspond to shorter hole drift times for the same interaction, and vice versa.)

To motivate our need to derive more complicated fitting functions we have fit our combined full spectra with a simple Gaussian shape function. As is evident in Fig.~\ref{fig:f3}, a simple Gaussian is not an adequate fit to the data, and specifically does not reproduce the underlying parameters we are trying to ascertain. Table~\ref{tab:662gaussian} gives the best-fit parameters to $661.66\,keV$ lines generated by our simulations for a range of drift times characteristic of our germanium detectors. Several trends in this table are worth noting for later comparison. First, the high $\chi_{R}^{2}$ values ($\gg\,1$) indicate poor quality fits, and get worse for larger drift times (i.e., more charge sharing). The fit value of $E_0$ is shifted to lower energies and $\sigma$ is broader than the actual noise value due to the Gaussian shape function trying to account for the low-energy tail. Utilizing this simple Gaussian shape function in spectral analyses can lead to erroneous gain calibrations as well as inaccurate measurement of peak energies and potential Doppler shifts and broadening for measured $\gamma$-ray lines. In addition, the simple Gaussian fit significantly overestimates the number of counts in the true photopeak ($N_p$) and leaves no characterization of the number of counts in the tail ($N_t$).

\section{Previous Analytical Fits}
\label{sect:previous}

A wide variety of shape functions have been proposed to facilitate analytical peak-fitting techniques for narrow $\gamma$--ray spectral lines in germanium detectors. These shape functions universally share the fundamental feature that the photopeak for totally-absorbed $\gamma$--rays where the resulting charge clouds are fully collected on the electrode(s) is modeled by a Gaussian peak of width $\sigma$, the center of which reflects the incident photon energy, $E_0$, in a properly calibrated system. The width of the Gaussian is determined by the statistical fluctuations in the initial e-h charge cloud produced by the recoil electron combined with the electronic readout noise \cite{jorch1977analytic}.

\begin{figure}
\centering
\includegraphics[width=1.0\textwidth]{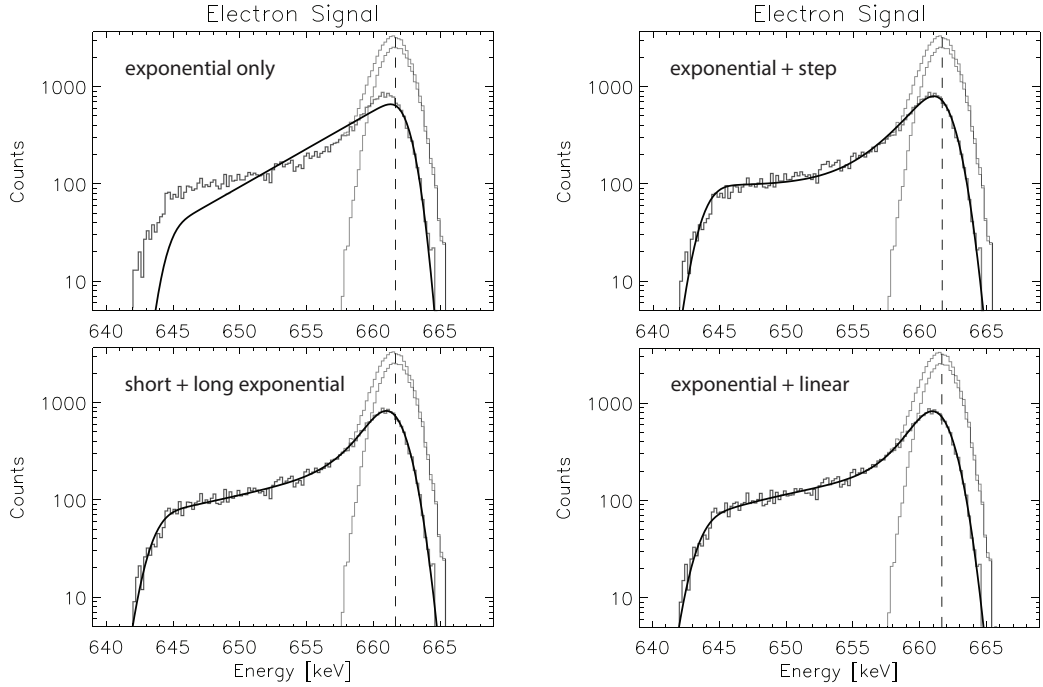}
\caption{\label{fig:f4} Example fits to tail-only spectra, 661.66 keV, $250\,ns$ drift time. The low-energy tail spectra were fit with the tail components of the shape function only, with all parameters unconstrained. Here we show only electron signal spectra, but the hole signal spectra and resulting fits are very similar. (Top left) Exponential tail only fit ($\chi_{R}^{2}$ = 13.37). (Top right) Exponential + step tail ($\chi_{R}^{2}$ = 1.98). (Bottom left) Short + long exponentials ($\chi_{R}^{2}$ = 1.27). (Bottom right) Exponential + linear tail ($\chi_{R}^{2}$ = 1.23). Both of the latter two shapes produce high quality fits and reproduce the photopeak energy and the number of counts in the tail; however, the last shape, exponential + linear tail, provides a much more robust fitting as we vary the interaction energy and drift time. The true photopeak and full spectra are shown only for comparison.}
\end{figure}

Measured spectral lines in germanium detectors always exhibit asymmetries. There is inevitably an excess of counts on the lower-energy side of the peak, low-energy tails, that have traditionally been attributed to a number of physical process in the detector including charge trapping, inactive regions in the detector, and escaped bremmstrahlung photons \cite{longoria1990analytical}. With the advent of multi-electrode detectors such as the COSI cross-strip detectors, charge sharing between multiple electrodes can be added to this list as a dominating contributor to low-energy tails. Occasionally spectral peaks exhibit excess events on the higher-energy side of the peak. Such high-energy tails are primarily due to electronic pile-up \cite{longoria1990analytical} or cross-talk between neighboring electrode electronics \cite{sleator2019benchmarking}. We will not consider high-energy tails further in this work. While charge trapping is present in our germanium detectors for both electron and hole signals, we are not including the effects of trapping on our simulated line profiles. We will return to a discussion of charge trapping in Section~\ref{sect:disc}.

Comparison of the wide variety of proposed shape functions for germanium detectors have been reviewed by multiple authors, e.g. \cite{mcnelles1975analytic,helmer1980analytical}. In general, the shape functions that best fit experimental data combine the Gaussian peak with an exponential low-energy tail, plus an additional component extending to lower energies that is usually represented by either a step function or a second longer exponential tail (or both) \cite{phillips1976automatic,varnell1969peak}. Given the simplicity and historical success of this general shape function, we adopt this as our baseline approach. 

These previous investigations into optimal shape functions were primarily modeling the response of monolithic, single-electrode germanium detectors, where the low-energy tails would extend indefinitely below the Gaussian peak. The low-energy tails we are simulating in this work are due solely to untriggered charge sharing on neighboring electrode strips, hence these low-energy tails only extend below the peak ($E_0$) to energies $E_0$-$E_{th}$, where $E_{th}$ is the trigger threshold energy for the neighboring strip. For our germanium cross strip detectors this threshold is at relatively low energies ($E_{th} = 18\,keV$). Hence we introduce our first modification to any adopted shape function by requiring a low-energy cutoff at $E_0$-$E_{th}$. 

We explored four (4) tailing shape functions in detail, keeping in mind our simultaneous goals of finding a shape function that provides a robust and high quality fit to the simulated spectra, reliably reproduces $E_0$, $\sigma$, $N_p$, and $N_t$, and minimizes the number of fit parameters. Here we define ``robust'' fits as ones where the parameters do not vary dramatically as we vary the interaction energy and the drift times. The four models we have explored are shown in  Fig.~\ref{fig:f4}, and summarized here. 

1. Exponential low-energy tail (2 parameters). Before adding any additional components to the tail shape function we first looked at the single-component exponential low-energy tail model. As can be seen in Fig.~\ref{fig:f4} (top left), this single component model does not provide a quality fit of the tail, justifying the need to look for an additional component to extend the tail to lower energies. 

2. Exponential + step tail (3 parameters). The simplest additional component we can add to this tail model is a step function that extends the tail shape function to lower energies. The form of this shape function is the same as used in \cite{pike2022properties}, but with the addition of the low-energy cutoff. An example fit utilizing this tail shape is shown in Fig.~\ref{fig:f4} (top right). This function does a much better job qualitatively of modeling the complex tail shape, but the quality of the fits ($\chi_{R}^{2}$) indicate there is room for further improvement, and close inspection shows that the fit does not adequately capture the slope of the extended tail at the lowest energies. Surprisingly, however, the fits with this tail shape are very robust and do an excellent job of reproducing the underlying physical parameters ($E_0$, $\sigma$, $N_p$, $N_t$), with minimal parameters. We keep this quality in mind when investigating the next two models. 

3. Short + long exponential tail (4 parameters). The next modification to the shape function we pursued replaces the step function with a second, longer exponential component to the tail \cite{varnell1969peak}. An example fit utilizing this double exponential is shown in Fig.~\ref{fig:f4} (bottom left). This shape function does an excellent job of producing high quality fits ($\chi_{R}^{2} \sim 1$) to the tail and overall full spectrum. That boded well for this model. However, we find that this tail shape function produces less robust fits (in terms of variation of the fitting parameters) than the exponential + step function, and also does an inferior job of accurately reproducing $E_0$, $\sigma$, $N_p$, and $N_t$.

4. Exponential + linear tail (4 parameters). The fourth option we explored is not as well represented in previous literature. We combined the short exponential tail with a linear function extending to lower energies. A fit utilizing this function is shown in Fig.~\ref{fig:f4} (bottom right). This shape function does an excellent job in fitting the extended tails, with $\chi_{R}^{2} \sim 1$, comparable to the double exponential function. However, this shape function also provides robust fits and reliably reproduces the underlying physical parameters ($E_0$, $\sigma$, $N_p$, $N_t$). Hence, we have selected this shape function as the most promising to pursue in greater detail.

\section{Empirical Tailing Model}
\label{sect:tail-fit}

We arrive at a shape function that includes three core components: a Gaussian peak (3 parameters:  $A$, $E_0$, $\sigma$), a short exponential low-energy tail (2 parameters: $B$, $\Gamma_S$), and a long linear low-energy tail (2 parameters: $C$, $D$). The latter two components need to be cut off at higher energies ($E_0$) and lower energies ($E_{0}-E_{th}$), as well as effectively broadened by a noise term (1 parameter, $\sigma_{t}$). This latter noise term for the tail component, $\sigma_{t}$, is often assumed equal to the noise terms in the Gaussian peak, $\sigma$, but not always \cite{mcnelles1975analytic,phillips1976automatic}. We have chosen to keep this as a free parameter for now and check whether consistency between these fitted parameters justifies setting them equal or not. Technically, the low-energy cutoff introduces an additional parameter to these fits, but this is a known parameter for our detectors and is fixed in these fits ($E_{th} = 18\,keV$). 

The shape function we have selected that reflects all of these components is given by the equation (8 parameters):

\begin{equation}
f(E) = Ae^{\frac{-(E-E_0)^2}{2\sigma^2}} + [Be^{\Gamma(E-E_0)} +C(1+D(E-E_0))]*[1-erf(\frac{E-E_0}{\sqrt{2}\sigma_{t}})]*[1+erf(\frac{E-E_0+E_{th}}{\sqrt{2}\sigma_{t}})]
\label{eqn:shapefunction}
\end{equation}

\noindent The last two terms in brackets represent the high- and low-energy cutoffs to the two tail components. 

\begin{figure}
\centering
\includegraphics[width=1.0\textwidth]{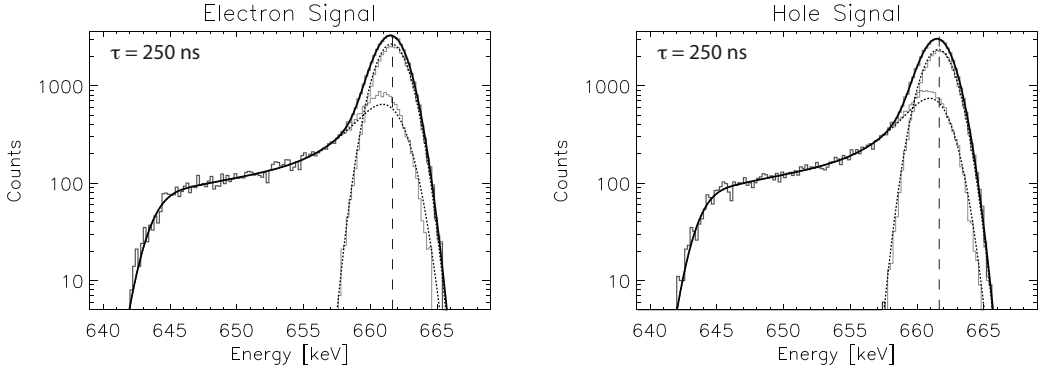}
\caption{\label{fig:f5} Example, $\tau = 250\,ns$, $E_0 = 661.66\,keV$, fits of the full shape function to full spectra (8 unconstrained parameters). The overall quality of fit is excellent: (Left) electron signals, ($\chi_{R}^{2} = 1.218$), (Right) hole signals, ($\chi_{R}^{2} = 1.104$). But as can seen in both plots the full shape function fit shifts and broadens the Gaussian photopeak while overestimating the number of counts in the true photopeak and underestimating the number of counts in the low-energy tail. The true photopeak and low-energy tail spectral components are shown for comparison with the relevant components of the fitted shape function (dotted lines) to illustrate how the unconstrained full shape function can incorrectly reflect the underlying components of the spectra.}
\end{figure}

In Fig.~\ref{fig:f4} (bottom right) we show just the tail components of this shape function (short exponential tail and longer linear tail) fit to the low-energy tail model spectra, keeping all of the fit parameters free. The tail shape function does an excellent job of recreating the profile of the simulated tail. It also does an excellent job of reproducing the true physical parameters that we are trying to uncover: the initial interaction energy, $E_0$, and the number of counts in the low-energy tail, $N_t$. 

However, we encounter a challenge when we try to fit the full shape function (Eqn.~\ref{eqn:shapefunction}) to the full spectral lines. As can be seen in Fig.~\ref{fig:f5}, this fit with eight unconstrained parameters does not adequately distinguish between the low-energy tail events and the true photopeak events. The result is that the fitted Gaussian peak is broader than the true photopeak, with the fitted peak energy shifted to lower energies. The number of counts in the true photopeak ($N_p$) are overestimated, while the number of counts in the low-energy tail ($N_t$) are underestimated. Effectively this full shape function fit is utilizing the eight (8) unconstrained fitting parameters to maximize the quality of the empirical fit (i.e., minimize $\chi_R^2$) at the expense of producing inaccurate physical numbers. While the quality of fit is promising, this model as implemented with eight unconstrained parameters does not meet our requirement that the shape function reliably reproduce the underlying physical parameters when fit to the full spectrum.

\section{Constrained Tailing Model}
\label{sect:constrain}

To address this challenge, we returned to our simulated low-energy tail spectra and the tail shape function to see if there are modifications we can make to the tail shape function to more robustly reproduce the actual tail parameters when doing the full shape function fit to the full spectrum. Here is where our ability to model the low-energy tail spectra separately from the true photopeak spectra becomes particularly powerful. In Table~\ref{tab:662fits} we show the best-fit parameters for the tail model fit to the low-energy tail spectra for a range of drift times characteristic of our germanium detectors. For these fits we held the fit parameter $E_{0}$ fixed at the known photon interaction energy (661.66\,keV) since this parameter is primarily driven in the full shape function fits by the Gaussian peak. When we look closely at these fit parameters three significant trends pop out.

\begin{table}
\centering
\begin{tabular}{ccccccc}
\hline
Electron signals\\
\hline
$\tau$ [ns] & $\Gamma$ [$keV^{-1}$] & $C/B$ & $D$ [$keV^{-1}$] & $\sigma_{t}/\sigma$ & $N_{t} [fit/sim]$ & $\chi_{R}^{2}$\\
\hline
10 & 0.53 & 0.14 & 0.030 & 0.81 & 0.98 & 0.96\\
50 & 0.50 & 0.13 & 0.027 & 0.85 & 0.99 & 0.92\\
100 & 0.50 & 0.13 & 0.027 & 0.86 & 0.99 & 1.05\\
150 & 0.50 & 0.13 & 0.028 & 0.86 & 0.99 & 1.40\\
200 & 0.51 & 0.13 & 0.029 & 0.85 & 0.99 & 1.23\\
250 & 0.51 & 0.13 & 0.029 & 0.85 & 0.99 & 1.20\\
\hline
Ave & 0.50 & 0.13 & 0.029 & 0.85\\
\hline
& & & \\
\hline
Hole signals\\
\hline
$\tau$ [ns] & $\Gamma$ [$keV^{-1}$] & $C/B$ & $D$ [$keV^{-1}$] & $\sigma_{t}/\sigma$ & $N_{t} [fit/sim]$ & $\chi_{R}^{2}$\\
\hline
10 & 0.51 & 0.15 & 0.030 & 0.85 & 0.98 & 0.87\\
50 & 0.49 & 0.13 & 0.028 & 0.86 & 0.99 & 1.51\\
100 & 0.50 & 0.13 & 0.029 & 0.85 & 0.99 & 1.45\\
150 & 0.51 & 0.14 & 0.030 & 0.85 & 0.99 & 1.31\\
200 & 0.50 & 0.13 & 0.029 & 0.85 & 0.99 & 1.29\\
250 & 0.51 & 0.13 & 0.030 & 0.84 & 0.99 & 1.33\\
\hline
Ave & 0.50 & 0.13 & 0.029 & 0.85\\
\hline

\end{tabular}
\caption{\label{tab:662fits} Best-fit parameters to the tail shape function holding $E_{0}$ fixed at 661.66\,keV but allowing all the other parameters to vary.}

\end{table}

First, the parameter $\Gamma$ in the fit for the exponential component of the low-energy tail does not vary significantly over the full range of drift times, nor between electron signals versus hole signals.  The limited range of these best-fit values suggests that we can help stabilize the full shape function fits by fixing $\Gamma$ for our tail shape function. We chose a value of $\Gamma$ averaged over both the electron signals and the hole signals as well as the range of drift times, weighted by the number of counts in the low-energy tails. This averaging resulted in us fixing the parameter $\Gamma \equiv 0.50\,keV^{-1}$ (at $E_{0} = 661.66\,keV$).

\begin{figure}
\centering
\includegraphics[width=1.0\textwidth]{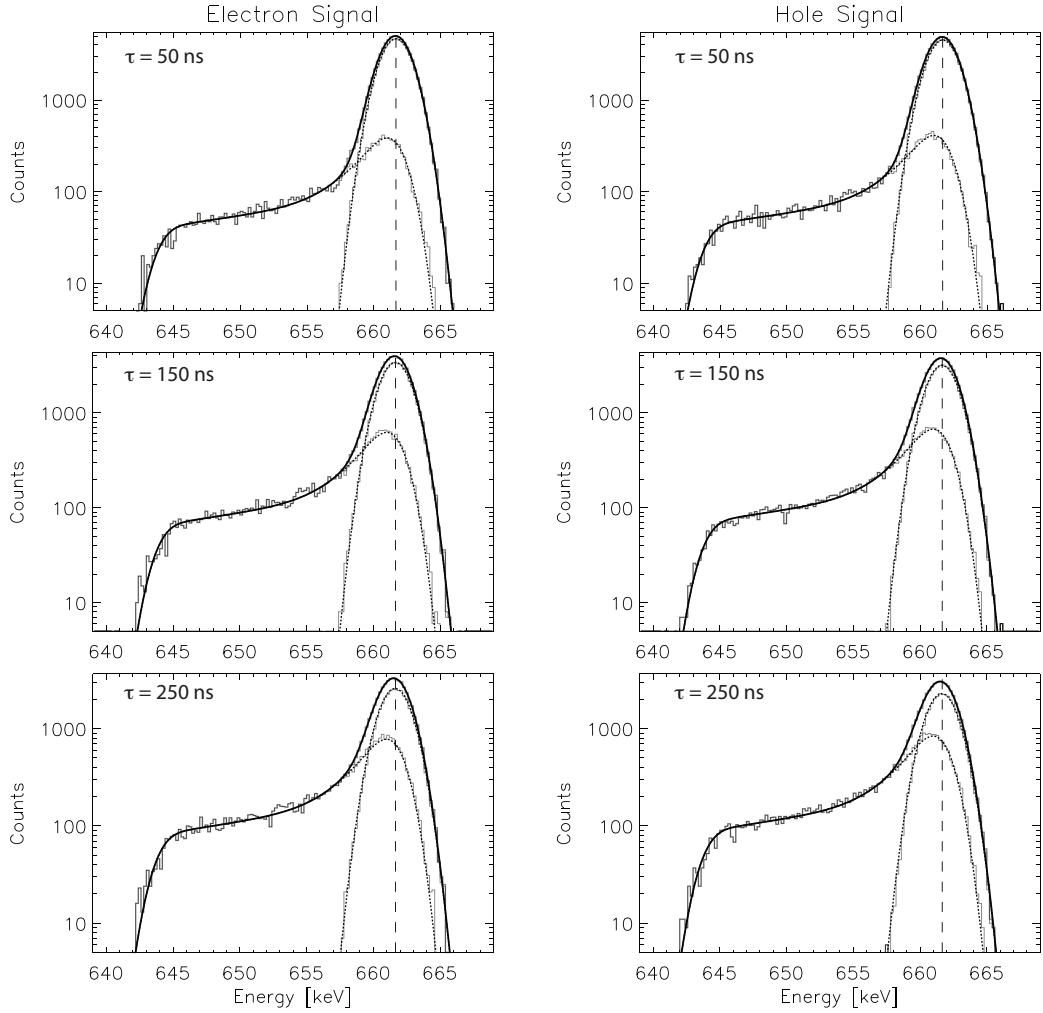}
\caption{\label{fig:f6}  Constrained fits to the full spectra, for three different drift times. (Left) Electron signals, (Right) hole signals. The quality of fits and corresponding fit parameters are listed in Table~\ref{tab:662constrained}. The individual true photopeak and low-energy tail components of both the model spectra and the fitted shape function are shown for comparison but were not used in the fitting.}
\end{figure}

The second trend that we can see in Table~\ref{tab:662fits} is that the ratio of the amplitudes of the linear tail to the exponential tail, $C/B$, also remains nearly consistent over the range of drift times for both the electron signals and the hole signals. This is our second clue to stabilizing the tail shape function fits by fixing the ratio $C/B$. Based again on the weighted average, we fixed the ratio $C/B \equiv 0.13$ (at $E_{0} = 661.66\,keV$).

The third trend that we can see in Table~\ref{tab:662fits} is that the slope of the linear component of the tail, $D$, also remains nearly consistent over the range of drift times for both the electron signals and the hole signals. This is our third clue to stabilizing the tail shape function fits by fixing the slope $D$. Based again on the weighted average, we fixed the slope $D \equiv 0.29\,keV^{-1}$ (at $E_{0} = 661.66\,keV$).

Finally, the fourth trend that we can see in Table~\ref{tab:662fits} is that the ratio of the noise terms, $\sigma_{t}/\sigma$, also remains nearly constant over the range of drift times for both the electron signals and the hole signals. This provides our fourth clue for stabilizing the tail shape function fits by fixing the ratio of these noise terms. Based again on the weighted average, we fixed the ratio $\sigma_{t}/\sigma \equiv 0.85$. Notably, this ratio is not unity, which justifies defining $\sigma_{t}$ as a separate parameter from $\sigma$. 

\begin{table}
\centering
\begin{tabular}{cccccc}
\hline
Electron signals\\
\hline
$\tau$ [ns] & $E_{0}$ [$keV$] & $\sigma$ [$keV$] & $N_{p} [fit/sim]$ & $N_{t} [fit/sim]$  & $\chi_{R}^{2}$\\
\hline
10 & 661.66 & 1.15 & 1.00 & 1.00 & 1.20\\
50 & 661.66 & 1.15 & 1.00 & 0.99 & 0.98\\
100 & 661.66 & 1.15 & 1.00 & 0.98 & 1.33\\
150 & 661.65 & 1.15 & 1.01 & 0.97 & 1.24\\
200 & 661.65 & 1.15 & 1.01 & 0.97 & 1.31\\
250 & 661.64 & 1.16 & 1.02 & 0.97 & 1.39\\
\hline
& & & \\

\hline
Hole signals\\
\hline
$\tau$ [ns] & $E_{0}$ [$keV$] & $\sigma$ [$keV$] & $N_{p} [fit/sim]$ & $N_{t} [fit/sim]$  & $\chi_{R}^{2}$\\
\hline
10 & 661.66 & 1.15 & 1.00 & 0.99 & 0.92\\
50 & 661.66 & 1.15 & 1.00 & 0.99 & 0.96\\
100 & 661.65 & 1.15 & 1.00 & 0.99 & 0.86\\
150 & 661.65 & 1.15 & 1.00 & 0.98 & 1.02\\
200 & 661.64 & 1.16 & 1.01 & 0.98 & 1.17\\
250 & 661.63 & 1.16 & 1.02 & 0.97 & 1.18\\
\hline

\end{tabular}
\caption{\label{tab:662constrained} Best-fit parameters for constrained fits to the full spectra. The simulated spectra assume $E_{0} = 661.66\,keV$ and $\sigma = 1.15\,keV$. The full shape function with constrained parameters ($\Gamma$, $C/B$, $D$, $\sigma_t/\sigma$ held fixed) still produces high-quality fits with $\chi_{R}^2$ comparable to the unconstrained version, but reliably reproduces $E_0$, $\sigma$, $N_p$, and $N_t$. }
\end{table}

By fixing the parameters $\Gamma$, $C/B$, $D$, and $\sigma_{t}/\sigma$, we have effectively reduced the number of free parameters in our shape function from eight (8) to four (4), just one additional parameter over the pure Gaussian peak ($B$). This additional parameter is effectively the amplitude of the low-energy tail component.

In Fig.~\ref{fig:f6} we show simulated spectra ($E_{0} = 661.66\,keV$) fit to the refined shape function, Eqn.~\ref{eqn:shapefunction} with $\Gamma$, $C/B$, $D$, and $\sigma_{t}/\sigma$ held fixed. The constrained shape function still does an excellent job in reproducing the overall shape of the full spectrum. It also does a better job of characterizing the underlying true photopeak and low-energy tail spectra. Most importantly, this constrained shape function reliably reproduces the underlying physical parameters as documented in Table~\ref{tab:662constrained}. The quality of fits ($\chi_{R}^{2}$) remain good for the full range of drift times. The true photopeak parameters, $E_{0}$ and $\sigma$ are consistently reproduced in the fits, and the number of counts within the true photopeak ($N_{p}$) are reliably reproduced with $\leq 2\%$ systematic error, and the number of counts within the low-energy tail ($N_{t}$) with $\leq 3\%$ systematic error. So far, as verified at $E_{0} = 661.66\,keV$ at least, this constrained tail fit has met our goals for the shape function (quality of fit, reliable parameter estimates, minimal parameters). 

\begin{table}
\centering
\begin{tabular}{cccccc}
\hline
$E_{0}$ [keV] & $<\Gamma>$ [$keV^{-1}$] & $<C/B>$ & $<D>$ [$keV^{-1}$] & $<\sigma_{t}/\sigma>$ & $<\chi_{R}^{2}>$\\
\hline
59.54 & 0.55 & 0.15 & 0.025 & 0.85 & 1.21\\
122.06 & 0.55 & 0.14 & 0.028 & 0.85 & 1.17\\
356.02 & 0.52 & 0.14 & 0.029 & 0.85 & 1.17\\
511.00 & 0.51 & 0.13 & 0.029 & 0.85 & 1.25\\
661.66 & 0.50 & 0.13 & 0.029 & 0.85 & 1.26\\
898.04 & 0.50 & 0.13 & 0.028 & 0.84 & 1.21\\
1173.24 & 0.49 & 0.13 & 0.028 & 0.84 & 1.19\\
1274.53 & 0.48 & 0.13 & 0.028 & 0.85 & 1.16\\
1332.50 & 0.48 & 0.13 & 0.027 & 0.85 & 1.11\\
1674.73 & 0.47 & 0.13 & 0.027 & 0.85 & 1.16\\
1836.06 & 0.46 & 0.12 & 0.027 & 0.85 & 1.19\\
\hline

\end{tabular}
\caption{\label{tab:generalfits} Best-fit parameters at various interaction energies $E_{0}$ to the tail shape function, averaged over electrons and holes as well as drift times, holding $E_{0}$ fixed but allowing all the other parameters to vary. These spectral simulations assume a point-like initial interaction (see Section~\ref{sect:extended}). These parameters are also plotted in Fig.~\ref{fig:f8}.}

\end{table}

\section{Generalization to Other Energies}
\label{sect:general}

Our constrained shape function works well for fitting the spectrum of monoenergetic interactions at 661.66\,keV ($^{137}Cs$). The immediate question is whether the constrained shape function can adequately meet our requirements for fitting at other $\gamma$--ray energies. To answer this question we reproduced the analysis above for a range of additional energies, representing common laboratory calibration source monoenergetic line energies: 59.54\,keV ($^{241}Am$), 122.06\,keV ($^{57}Co$), 356.02\,keV ($^{133}Ba$), 511.00\,keV ($^{22}Na$), 898.04\,keV ($^{88}Y$), 1173.24\,keV ($^{60}Co$), 1274.54\,keV ($^{22}Na$), 1332.50\,keV ($^{60}Co$), 1674.73\,keV ($^{58}Co$), and 1836.06\,keV ($^{88}Y$). The primary factors that vary in the detector response for these energies are that the measured noise $\sigma_m$ increases with energy (Eqn.~\ref{eqn:noise}), and the effects of repulsion on the charge cloud profile are larger at higher energies \cite{boggs2023}. Table~\ref{tab:generalfits} documents the average best-fit parameters over this range of energies. The quality of fit ($<\chi_{R}^{2}>$) remains very good over the full range of energies. As at $661.66\,keV$, $\Gamma$, $C/B$, and $D$ do not vary significantly for a given interaction energy between electron signals and hole signals and as we varied the drift times, \emph{but the average values of these parameters do vary with photon energy itself.} The ratio $<\sigma_{t}/\sigma>$ varies very little with energy, remaining nearly constant at $<\sigma_{t}/\sigma> \sim 0.85$. 

\begin{figure}
\centering
\includegraphics[width=1.0\textwidth]{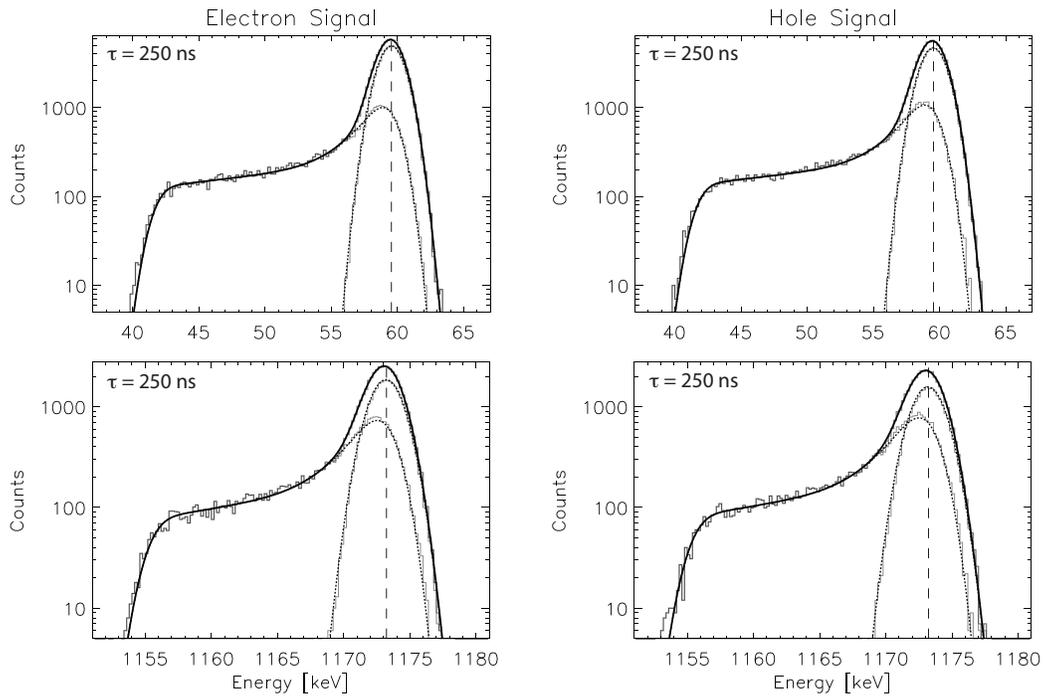}
\caption{\label{fig:f7} Example constrained shape function fits. (Top Left) 59.54\,keV, electron signals ($\chi_{R}^2 = 1.21$). (Top Right) 59.54\,keV, hole signals ($\chi_{R}^2 = 1.03$).  (Bottom Left) 1173.23\,keV, electron signals ($\chi_{R}^2 = 1.05$). (Bottom Right) 1173.23\,keV, hole signals ($\chi_{R}^2 =1.20$). $E_0$, $\sigma$, $N_p$, and $N_t$ remain well reproduced by the fits.}
\end{figure}

In Fig.~\ref{fig:f7} we show example fits for 59.54\,keV and 1173.23\,keV model spectra, for drift times of 250\,ns (maximum tailing), utilizing the energy-specfic fixed parameters from Table~\ref{tab:generalfits}. The quality of fits remain excellent despite the wide range in energy.

\section{Extended Initial Charge Cloud}
\label {sect:extended}

Now we turn to the impact of extended charge distributions of electron-hole pairs created by the recoil electron following the initial $\gamma$--ray interaction. Extended initial charge cloud distributions will become increasingly important for higher interaction energies, such as those we are measuring in $\gamma$--ray applications of our germanium detectors. In \citet{boggs2023}, we discussed a simple approach to modeling the effects of initially extended charge clouds by assuming that the initial charge cloud can be approximated at $t=0$ as a sphere of uniform charge density and finite radius $R_0$. The size of this initial charge cloud can be estimated using the practical electron range, $D_p$, in germanium as a function of recoil electron energy, which can be estimated by the following formula \cite{wohl1984review}:

\begin{equation}
D_{p}(E) = \alpha E[1-\beta /(1+\gamma E)]
\label{eqn:range}
\end{equation}

\noindent With $\alpha = 0.83\,\mu m \cdot keV^{-1}$, $\beta = 0.9841$, and $\gamma = 0.0030\,keV^{-1}$. This estimate of the practical range for electrons in germanium is accurate to $\sim 10\%$, which is adequate for our purposes. For our charge cloud estimates, we assume for a given $E_0$ that $R_0 = D_p/2$, effectively that the recoil electron deposits its energy uniformly in a sphere with a diameter equal to the full practical range. This assumption is conservative, overestimating the full extent of the initial charge cloud and hence the impacts of charge sharing on simulated tail spectra. 

\begin{figure}
\centering
\includegraphics[width=1.0\textwidth]{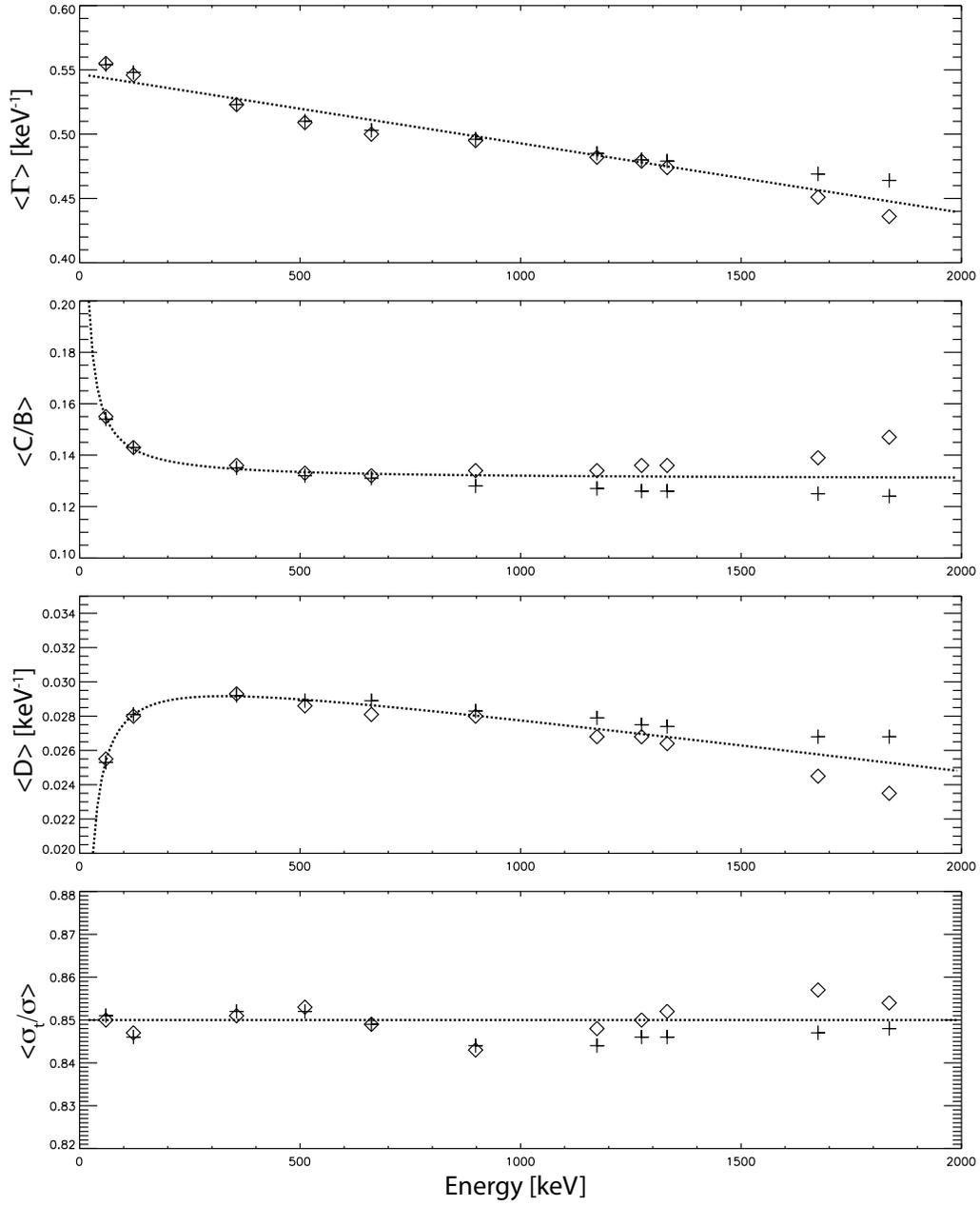}
\caption{\label{fig:f8} The best-fit average tail shape function parameters $<\Gamma>$, $<C/B>$, $<D>$, and $<\sigma_{t}/\sigma>$ for our range of interaction energies, showing the parameters derived for point-like initial interactions (crosses) and extended initial charge clouds (diamonds). Also shown are the best-fit curves (dotted lines) to $\Gamma(E)$ (Eqn.~\ref{eqn:gamma}), $C/B(E)$ (Eqn.~\ref{eqn:coverb}), and $D(E)$ (Eqn.~\ref{eqn:dvse}) for the average of these two extreme cases. }
\end{figure}

We have repeated the analysis performed above for the characteristic $\gamma$--ray calibration energies of fitting the low-energy tail spectra and determining the average best-fit parameters. These results are shown in Table~\ref{tab:extendedfits} and Fig.~\ref{fig:f8}. These results show that the adopted tail shape function continues to provide an excellent fit to the tail spectra, both in terms of quality of fit ($<\chi_{R}^{2}>$) as well as reproducing the physical parameters, even when the spectra include the effects of an extended initial charge cloud. Below interaction energies of $\sim 1000\,keV$ the recoil electron range has little effect on the fits. At higher energies, we can start to see slight divergence in the best-fit parameters for $<\Gamma>$, $<C/B>$, and $<D>$ compared to the point interaction case, though the ratio $<\sigma_{t}/\sigma>$ remains largely unchanged between the two cases. 

\begin{table}
\centering
\begin{tabular}{cccccc}
\hline
$E_{0}$ [keV] & $<\Gamma>$ [$keV^{-1}$] & $<C/B>$ & $<D>$ [$keV^{-1}$] & $<\sigma_{t}/\sigma>$ & $<\chi_{R}^{2}>$\\
\hline
59.54 & 0.56 & 0.16 & 0.026 & 0.85 & 1.25\\
122.06 & 0.55 & 0.14 & 0.028 & 0.85 & 1.18\\
356.02 & 0.52 & 0.14 & 0.029 & 0.85 & 1.23\\
511.00 & 0.51 & 0.13 & 0.029 & 0.85 & 1.23\\
661.66 & 0.50 & 0.13 & 0.028 & 0.85 & 1.26\\
898.04 & 0.50 & 0.13 & 0.028 & 0.84 & 1.20\\
1173.24 & 0.48 & 0.13 & 0.027 & 0.85 & 1.11\\
1274.53 & 0.48 & 0.14 & 0.027 & 0.85 & 1.21\\
1332.50 & 0.47 & 0.14 & 0.026 & 0.85 & 1.11\\
1674.73 & 0.45 & 0.14 & 0.025 & 0.86 & 1.30\\
1836.06 & 0.44 & 0.15 & 0.024 & 0.85 & 1.13\\
\hline
\end{tabular}
\caption{\label{tab:extendedfits} Extended initial charge cloud. Best-fit parameters at various interaction energies $E_{0}$ to the tail shape function, averaged over electrons and holes as well as drift times, holding $E_{0}$ fixed but allowing all the other parameters to vary. These spectral simulations assume a spherical extended initial charge cloud. These parameters are also plotted in Fig.~\ref{fig:f8}.}
\end{table}

Given that our initial analysis in Section~\ref{sect:general} ignored the recoil electron range by assuming the initial charge clouds were concentrated at a single point, and the analysis in this section overestimates the effects of extended initial charge clouds by assuming spherical symmetry, we anticipate that the best-fit parameters for a realistic initial charge cloud distribution would lie somewhere between these two extremes. As seen in Table~\ref{tab:generalfits} and Table~\ref{tab:extendedfits}, these parameters are identical below $\sim 1000\,keV$. For higher energies, we can average the results for the two extreme cases to give an estimate of the best-fit parameters for realistic initial charge cloud distributions. Curve fits to the average of these best-fit parameters are plotted in Fig.~\ref{fig:f8}. Fortunately, these parameters vary smoothly and predictably as a function of $\gamma$--ray interaction energy. The energy dependence ($[E] = keV$) of the parameter $\Gamma(E)$ is fit by the function:

\begin{equation}
\Gamma(E) = 0.547 - (5.39\times10^{-5})E; [\Gamma] = keV^{-1}
\label{eqn:gamma}
\end{equation}

\noindent The ratio $C/B(E)$ is fit by the function:

\begin{equation}
C/B(E) = 0.131 + \frac{1.44}{E}
\label{eqn:coverb}
\end{equation}

\noindent The parameter $D(E)$ is fit by the function:

\begin{equation}
D(E) = 0.0312 - \frac{0.336}{E} - (3.13\times10^{-6})E; [D] = keV^{-1}
\label{eqn:dvse}
\end{equation}

\noindent We can effectively fix $\sigma_{t}/\sigma \equiv 0.85$ for all energies. Characterization of these trends enable us to utilize the constrained tail shape function at arbitrary photopeak energies with some confidence that we are accounting for the finite range of the initial recoil electron. 

\section{Discussion}
\label {sect:disc}

In order to optimize the spectral performance of high resolution germanium detectors, very careful energy calibrations need to be performed using monoenergetic $\gamma$--ray line sources of known energies. As seen with the simple Gaussian fits to the true asymmetric line profiles presented in Section~\ref{sect:spectra}, utilizing ill-fitting shape functions to fit the asymmetric line profiles will lead to incorrect determination of the true photoabsorption peak, and hence skew the subsequent energy calibrations for the detector. Conversely, detailed analysis of scientific spectral data requires a detailed understanding of the instrumental line profiles to accurately identify line energies, as well as potential Doppler broadening and shifts -- important factors in our astrophysical program.  

In this work we have considered the effects of charge sharing as the dominant factor in creating low-energy tails in our multi-electrode germanium detectors. We have not included charge trapping in these spectral simulations. While charge trapping can be a significant factor in producing low-energy tails in germanium detectors, our ability to measure the full 3-D position of photon interactions within our detector volume allows us to largely correct the effects of trapping on the collected spectra. The details of the charge trapping correction are beyond the scope of this current work, but suffice it to say that our work to correct the effects of charge trapping largely led to our in-depth analysis of spectral profiles presented in this paper. 

One of the advantages of the work presented here is the simplicity of the simulations used to generate the charge-sharing spectra, as well as the simplicity of the shape function used to characterize the resulting photopeaks and low-energy tails. While the shape function utilized in this work produces complicated line profiles, these profiles reliably reproduce the underlying physical parameters of the simulated spectra with only one additional parameter over a simple Gaussian peak. 

The shape function developed in this work has multiple future applications to the COSI program. The line profile will enable us to accurately perform the energy calibration of the instrument, including characterizing and correcting the effects of charge trapping in the detectors. The profiles themselves provide a useful tool for simulating the expected spectral performance of the instrument. Finally, the shape function will be a critical component in the scientific analysis and interpretation of the astrophysical data.

\section{Acknowledgements}
\label{}

This work was supported by the NASA Astrophysics Research and Analysis (APRA) program, grant 80NSSC21K1815. Thanks to J. Tomsick for feedback on this work.

\bibliographystyle{elsarticle-num-names} 
\bibliography{refs}

\end{document}